\documentclass{aaV7.0}
\usepackage{longtable}
\usepackage{natbib}
\usepackage{graphicx}

\bibpunct[, ]{(}{)}{;}{a}{}{,}
\usepackage{color}

\begin{document}

\title{Abundances of neutron-capture elements in \object{G\,24-25}
\thanks{Based on observations made with the Nordic Optical Telescope
on La Palma. Table 2 is available in electronic form at
{\tt http://www.aanda.org}}}

\subtitle{A halo-population CH subgiant}

\author{
 S. Liu\inst{1,2}, P. E. Nissen\inst{3}, W. J. Schuster\inst{4}, G. Zhao\inst{1}, Y. Q. Chen\inst{1} and Y. C. Liang\inst{1}
}

%\offprints{G. Zhao; \email{gzhao@bao.ac.cn}}
\institute{
  Key Laboratory of Optical Astronomy, National Astronomical
  Observatories, CAS, 20A Datun Road, Chaoyang District, 100012, Beijing,
  China\\
\email{gzhao@nao.cas.cn}
\and Graduate University of the Chinese Academy of
  Sciences, 19A Yuquan Road, Shijingshan District, 100049, Beijing,
  China
\and Department of Physics and Astronomy, University of Aarhus,
DK--8000 Aarhus C, Denmark\\
\email{pen@phys.au.dk}
\and Observatorio Astron\'{o}mico
Nacional, Universidad Nacional Aut\'{o}noma de M\'{e}xico, Apartado
Postal 877, C.P. 22800 Ensenada, B.C., M\'{e}xico\\
\email{schuster@astrosen.unam.mx}
}

\date{Received 18 August 2011 / Accepted 5 March 2012}

\abstract{}{The differences between the neutron-capture element abundances
of halo stars
are important to our understanding of the nucleosynthesis of elements
heavier than the iron group. We present a detailed abundance analysis
of carbon and twelve
neutron-capture elements from Sr up to Pb
for a peculiar halo star \object{G\,24-25} with [Fe/H] = $-1.4$ in order to probe
its origin.}
{The equivalent widths of unblended lines are measured from  high resolution NOT/FIES spectra
and used to derive abundances based on Kurucz model atmospheres.
In the case of CH, Pr, Eu, Gd, and Pb lines, the abundances are derived by
fitting synthetic profiles to the observed spectra.
Abundance analyses are performed both relative to the Sun and
to a normal halo star \object{G\,16-20} that has similar stellar parameters
as \object{G\,24-25}.}
{We find that \object{G\,24-25} is a halo subgiant star  with an unseen component.
It has large overabundances of carbon and heavy $s$-process elements
and mild overabundances of Eu and light $s$-process elements. This abundance
distribution is consistent with that of a typical CH giant. The abundance pattern
can be explained by mass transfer from a former asymptotic giant branch component, which is now a white dwarf.} {}


\titlerunning{Abundances of neutron-capture elements in \object{G\,24-25}}
\authorrunning{S. Liu et al.}

\maketitle
% _______________________________________________________________
%
\section{Introduction}\label{Sect:intro}
The heaviest elements in halo stars are mainly produced by neutron captures
divided into the slow ($s$-) and  the rapid ($r$-) processes,
which have different reaction timescales. Among the $s$-process synthesized
elements, lead (Pb) is particularly interesting because it is the most abundant element
in the third peak of the nuclei distribution in stars.  However,
Pb lines are very weak in low metallicity stars, and thus Pb abundances have not
been measured for many stars compared with other $s$-process elements, such as Ba.
Pb is often detected for carbon-enhanced stars with enhanced Ba abundances.
For example, \citet{Bisterzo06} compiled a catalog of 23 Pb-rich stars found in the literature.
Most of these stars have [Fe/H] $< -2$ and are carbon-enhanced metal-poor stars
with $s$-process enhanced abundances (CEMP-$s$) or with $r$- and $s$-process enhanced abundances
(CEMP-$r/s$). Eleven stars of the CEMP-$s$/CEMP-$r/s$ compiled in Bisterzo's work are proven members
of binary systems \citep{MC90, Aoki01, aoki02, Luca03, Cohen03, Siva04}.

In the past ten years, some studies also reported Pb measurements for
Ba stars and CH giants and subgiants \citep{VanEck01, VanEck03, John04, AB06, Pere09, PD11, Goswa10}.These classes of stars exhibit overabundances of $s$-process elements; in addition, the CH stars have high C abundances.  Furthermore, these stars fall into different metallicity
ranges and belong to different populations in the Galaxy. Most of the Ba stars are
found in the disk with [Fe/H] from $-$0.5 to +0.4; fewer than 6$\%$ of the Ba stars belong
to the halo \citep{Menn97, AB06}.
The classical CH giants are clearly members of the halo population with $-1.5<$[Fe/H]$<-1.0$ \citep{Vant92a}.
The CH subgiants, i.e stars with similar abundance distributions as the CH giants, but situated on either
the main sequence or the subgiant branch, have typical [Fe/H] values in
the range from $-0.5$ to $-0.2$ and disk population kinematics \citep{LB82, LB91, Sneden83, Smith93}.

Systematic spectroscopic studies indicate that all Ba and CH stars belong to binary systems \citep{MC90}.
At low metallicity, [Fe/H]$<-1.8$, \citet{Luca05} found that 68$\%$ of 19 CEMP-$s$ stars display evidence of radial velocity variations. This high frequency suggests that all CEMP-s stars are metal-poor analogs of the classical CH giants. The origin of the enhanced carbon and $s$-process elements
(including Pb) in these stars could be explained by mass transfer from a former
asymptotic giant branch (AGB) companion
(now a white dwarf) in a binary system.

In a study of two distinct halo populations in the solar neighborhood, \citet{NS2010, NS2011}
measured the abundance ratios and kinematics of 94 dwarf and subgiant
stars and found a strong enhancement of [Ba/Fe] for a halo star,
\object{G\,24-25}, with [Fe/H] = $-1.4$. Further
investigation of the spectrum identified a Pb line at 4057.8\,\AA, which
is not seen for the rest of the stars. For example, the Pb feature at 4057.8\,\AA\
cannot be detected in \object{G\,16-20}, a halo star with a similar metallicity as \object{G\,24-25}.
In addition, \object{G\,24-25} has a strong CH band near 4300\,\AA ,
indicating that it is a CH subgiant. To our knowledge, this is the first known CH subgiant
with a metallicity that is typical of the halo population.
In this paper, we report our abundance measurements of Pb and other neutron-capture
elements for this peculiar halo star. For comparison, \object{G\,16-20} is analyzed as a
reference star. The abundance pattern is compared with other chemically peculiar stars
in the literature, and simple AGB wind-accretion models are adopted to reproduce
the measured abundances.

\section{Abundance analysis}\label{Sect:abundances}

\subsection{Observational data and atmospheric parameters}
\label{Sect:Observations}
As described in \citet{NS2010}, the spectra of \object{G\,24-25} and \object{G\,16-20} were obtained at the 2.56m Nordic Optical Telescope (NOT) using the FIbre fed Echelle Spectrograph (FIES) with a resolving power of R $\sim$ 40,000. The spectral range goes from 4000\,\AA\ to 7000\,\AA ,
and the signal-to-noise ratio ($S/N$) equals about 170 at 5500\,\AA\  but is
only between 50 and 80 in the blue spectral region.

The stellar atmospheric parameters for \object{G\,24-25} and  \object{G\,16-20} are taken from \citet{NS2010} and given in Table~\ref{Tab:StellarParameters}, where the heliocentric radial velocities are also given. According to \citet{NS2010}, the total space velocities with respect to the local standard of rest are 315\,km\,s$^{-1}$ and 263\,km\,s$^{-1}$
for \object{G\,24-25} and  \object{G\,16-20}, respectively, which clearly classify both stars as members of the halo population.

\begin{table}
 \centering
 \caption{\label{Tab:StellarParameters} Atmospheric parameters and radial velocities of \object{G\,24-25} and \object{G\,16-20}.}
  \begin{tabular}{cccccc}
   \hline\hline
   \noalign{\smallskip}
   Star & $T_{\mbox{\scriptsize eff}}$  & $\log g$ & $\mathrm{[Fe/H]}$ & $\xi$ & $v_{\mbox{\scriptsize rad}}$  \\
        &    (K)  &  &   & (km~s$^{-1}$)  &  (km~s$^{-1}$)  \\
   \hline
   \noalign{\smallskip}
   \object{G\,24-25} & $5828$ & $3.86$ & $-1.40$ & $1.20$ & $-312.9$ \\
   \object{G\,16-20} & $5625$ & $3.64$ & $-1.42$ & $1.50$ & $~~170.8$ \\
   \noalign{\smallskip}
   \hline
  \end{tabular}
\end{table}

%\longtab{2}{
\onltab{2}{
\clearpage \onecolumn
\begin{longtable} {crcrrrrcrr}
 \caption{\label{Tab:EWandloggf} Atomic data and equivalent widths of spectral lines, as well as the derived abundances for \object{G\,24-25} and \object{G\,16-20}.\\
 This table contains the following information: wavelength in
 angstroms (Column 1), element identification (Column 2),
 excitation potential (Column 3), log of the oscillator strength (Column 4),
 reference for log{\it\,gf} (Column 5), measured equivalent width
 in milli-angstroms and derived abundance of {\object{G\,24-25}} (Columns 6 and 7),
 measured equivalent width in milli-angstroms and derived abundance of {\object{G\,16-20}} (Columns 8 and 9).
 }\\
\hline\hline
\noalign{\smallskip}
    Wavelength & Elem. & E. P. & log{\it\,gf} & Ref. & \multicolumn{2}{c}{\object{G\,24-25}} & & \multicolumn{2}{c}{\object{G\,16-20}}\\
   \cline{6-7}
   \cline{9-10}
   \noalign{\smallskip}
     (\AA)  &     & (eV)   &   &   & $EW$~(m\AA) & log~$\varepsilon$ &  & $EW$~(m\AA) & log~$\varepsilon$\\
%   \noalign{\smallskip}
\endfirsthead
\caption{continued.}\\
\hline\hline
\noalign{\smallskip}
    Wavelength & Elem. & E. P. & log{\it\,gf} & Ref. & \multicolumn{2}{c}{\object{G\,24-25}} & & \multicolumn{2}{c}{\object{G\,16-20}}\\
   \cline{6-7}
   \cline{9-10}
   \noalign{\smallskip}
     (\AA)  &     & (eV)   &   &   & $EW$~(m\AA) & log~$\varepsilon$ &  & $EW$~(m\AA) & log~$\varepsilon$\\
    \hline
    \noalign{\smallskip}
\endhead
\hline
\endfoot
\hline
\noalign{\smallskip}
   4932.049 & \ion{C}{i}  & 7.68 & $-1.68$ &  1 &  14.6 & 8.08 &&   ... &  ...\\
   5052.167 & \ion{C}{i}  & 7.68 & $-1.30$ &  1 &  23.8 & 7.97 &&   1.2: & 6.58:\\
   5380.337 & \ion{C}{i}  & 7.68 & $-1.62$ &  1 &  14.1 & 8.02 &&   ... &  ...\\
            &             &      & $     $ &    &       &      &&       &     \\
   4607.340 & \ion{Sr}{i} & 0.00 & $ 0.28$ &  2 &  18.0 & 1.96 &&   5.7 & 1.19\\
   4077.714 & \ion{Sr}{ii}& 0.00 & $ 0.17$ &  2 & 234.8 & 1.79 && 154.3 & 1.09\\
            &             &      & $     $ &    &       &      &&       &     \\
   4398.010 & \ion{Y}{ii} & 0.13 & $-1.00$ &  3 &  50.9 & 1.58 &&  18.3 & 0.56\\
   4883.690 & \ion{Y}{ii} & 1.08 & $ 0.07$ &  3 &  56.1 & 1.52 &&  25.0 & 0.60\\
   4900.110 & \ion{Y}{ii} & 1.03 & $-0.09$ &  3 &  56.2 & 1.63 &&  23.5 & 0.67\\
   5087.430 & \ion{Y}{ii} & 1.08 & $-0.17$ &  3 &  43.4 & 1.45 &&  15.0 & 0.54\\
   5123.220 & \ion{Y}{ii} & 0.99 & $-0.83$ &  3 &  22.0 & 1.51 &&   4.5 & 0.53\\
   5200.420 & \ion{Y}{ii} & 0.99 & $-0.57$ &  3 &  30.8 & 1.46 &&   8.8 & 0.58\\
   5205.730 & \ion{Y}{ii} & 1.03 & $-0.34$ &  3 &  39.3 & 1.47 &&  14.7 & 0.64\\
   5402.780 & \ion{Y}{ii} & 1.84 & $-0.44$ &  3 &   8.6 & 1.44 &&   ... &  ...\\
            &             &      & $     $ &    &       &      &&       &     \\
   4687.800 & \ion{Zr}{i} & 0.73 & $ 0.55$ &  4 &   4.7 & 2.14 &&   ... &  ...\\
   4050.330 & \ion{Zr}{ii}& 0.71 & $-1.06$ &  5 &  23.2 & 2.06 &&  11.5 & 1.47\\
   4090.510 & \ion{Zr}{ii}& 0.76 & $-1.01$ &  6 &  26.7 & 2.14 &&   ... &  ...\\
   4161.200 & \ion{Zr}{ii}& 0.71 & $-0.59$ &  5 &  46.7 & 2.15 &&  24.1 & 1.39\\
   4208.990 & \ion{Zr}{ii}& 0.71 & $-0.51$ &  5 &  47.0 & 2.07 &&  26.1 & 1.36\\
   4258.050 & \ion{Zr}{ii}& 0.56 & $-1.20$ &  5 &  27.2 & 2.13 &&   ... &  ...\\
   5350.080 & \ion{Zr}{ii}& 1.83 & $-1.24$ &  6 &   2.9 & 2.20 &&   ... &  ...\\
   5350.310 & \ion{Zr}{ii}& 1.77 & $-1.16$ &  5 &   4.2 & 2.22 &&   ... &  ...\\
            &             &      & $     $ &    &       &      &&       &     \\
   5853.688 & \ion{Ba}{ii}& 0.60 & $-0.91$ &  7 &  94.6 & 2.15 &&  29.5 & 0.55\\
   6141.727 & \ion{Ba}{ii}& 0.70 & $-0.03$ &  7 & 163.1 & 2.08 &&  69.0 & 0.53\\
   6496.908 & \ion{Ba}{ii}& 0.60 & $-0.41$ &  7 & 142.5 & 2.13 &&  61.6 & 0.64\\
            &             &      & $     $ &    &       &      &&       &     \\
   4526.110 & \ion{La}{ii}& 0.77 & $-0.59$ &  8 &  15.0 & 1.22 &&   ... &  ...\\
   4662.510 & \ion{La}{ii}& 0.00 & $-1.24$ &  8 &  22.8 & 1.28 &&   ... &  ...\\
   4748.730 & \ion{La}{ii}& 0.93 & $-0.54$ &  8 &  14.7 & 1.23 &&   ... &  ...\\
   4920.960 & \ion{La}{ii}& 0.13 & $-0.58$ &  8 &  48.8 & 1.38 &&  10.1 & 0.06\\
   4921.780 & \ion{La}{ii}& 0.24 & $-0.45$ &  8 &  45.2 & 1.27 &&  10.6 & 0.07\\
            &             &      & $     $ &    &       &      &&       &     \\
   4053.490 & \ion{Ce}{ii}& 0.00 & $-0.61$ &  9 &  29.5 & 1.71 &&   ... &  ...\\
   4062.230 & \ion{Ce}{ii}& 1.37 & $ 0.30$ & 10 &  19.6 & 1.88 &&   1.2: & 0.36:\\
   4083.220 & \ion{Ce}{ii}& 0.70 & $ 0.27$ &  9 &  36.1 & 1.70 &&   ... &  ...\\
   4117.290 & \ion{Ce}{ii}& 0.74 & $-0.45$ &  9 &  11.3 & 1.69 &&   ... &  ...\\
   4118.140 & \ion{Ce}{ii}& 0.70 & $ 0.13$ &  9 &  32.4 & 1.74 &&   ... &  ...\\
   4120.830 & \ion{Ce}{ii}& 0.32 & $-0.37$ &  9 &  31.9 & 1.85 &&   ... &  ...\\
   4127.360 & \ion{Ce}{ii}& 0.68 & $ 0.31$ &  9 &  40.9 & 1.76 &&   ... &  ...\\
   4153.130 & \ion{Ce}{ii}& 0.23 & $-0.80$ &  9 &  16.7 & 1.74 &&   ... &  ...\\
   4222.600 & \ion{Ce}{ii}& 0.12 & $-0.15$ &  9 &  46.6 & 1.80 &&   6.6 & 0.28\\
   4427.920 & \ion{Ce}{ii}& 0.54 & $-0.41$ &  9 &  19.2 & 1.71 &&   ... &  ...\\
   4460.230 & \ion{Ce}{ii}& 0.47 & $ 0.28$ & 10 &   ... &  ... &&  10.7 & 0.41\\
   4483.900 & \ion{Ce}{ii}& 0.86 & $ 0.10$ &  9 &  29.0 & 1.79 &&   ... &  ...\\
   4523.080 & \ion{Ce}{ii}& 0.52 & $-0.08$ &  9 &   ... &  ... &&   3.1 & 0.24\\
   4539.740 & \ion{Ce}{ii}& 0.33 & $-0.08$ &  9 &  42.5 & 1.78 &&   5.4 & 0.30\\
   4560.960 & \ion{Ce}{ii}& 0.68 & $-0.26$ &  9 &  23.5 & 1.82 &&   ... &  ...\\
   4562.360 & \ion{Ce}{ii}& 0.48 & $ 0.21$ &  9 &  50.6 & 1.85 &&   7.0 & 0.29\\
   4565.840 & \ion{Ce}{ii}& 1.09 & $ 0.07$ &  9 &  18.2 & 1.74 &&   ... &  ...\\
   4628.160 & \ion{Ce}{ii}& 0.52 & $ 0.14$ &  9 &  46.1 & 1.83 &&   ... &  ...\\
   5044.030 & \ion{Ce}{ii}& 1.21 & $-0.14$ &  9 &  10.4 & 1.72 &&   ... &  ...\\
   5187.450 & \ion{Ce}{ii}& 1.21 & $ 0.17$ &  9 &  18.1 & 1.71 &&   ... &  ...\\
   5274.230 & \ion{Ce}{ii}& 1.04 & $ 0.13$ &  9 &  23.3 & 1.72 &&   ... &  ...\\
   5330.540 & \ion{Ce}{ii}& 0.87 & $-0.40$ &  9 &  12.4 & 1.72 &&   ... &  ...\\
   5468.380 & \ion{Ce}{ii}& 1.40 & $-0.07$ &  9 &   8.9 & 1.75 &&   ... &  ...\\
   5610.246 & \ion{Ce}{ii}& 1.05 & $-0.74$ & 10 &   6.0 & 1.85 &&   ... &  ...\\
            &             &      & $     $ &    &       &      &&       &     \\
   5322.800 & \ion{Pr}{ii}& 0.48 & $-0.32$ & 11 &   syn & 1.28 &&   syn & $<-0.38$\\
            &             &      & $     $ &    &       &      &&       &     \\
   4012.700 & \ion{Nd}{ii}& 0.00 & $-0.74$ & 12 &  27.0 & 1.56 &&   4.3 & 0.35\\
   4051.150 & \ion{Nd}{ii}& 0.38 & $-0.30$ & 13 &  28.5 & 1.54 &&   6.4 & 0.48\\
   4232.380 & \ion{Nd}{ii}& 0.06 & $-0.47$ & 13 &  35.1 & 1.53 &&   ... &  ...\\
   4542.603 & \ion{Nd}{ii}& 0.74 & $-0.28$ & 13 &  17.2 & 1.49 &&   2.4 & 0.34\\
   4645.770 & \ion{Nd}{ii}& 0.56 & $-0.76$ & 13 &  10.0 & 1.49 &&   ... &  ...\\
   5089.830 & \ion{Nd}{ii}& 0.28 & $-1.40$ &  2 &   5.5 & 1.53 &&   ... &  ...\\
   5092.780 & \ion{Nd}{ii}& 0.30 & $-0.61$ & 13 &  21.4 & 1.46 &&   2.6 & 0.29\\
   5192.620 & \ion{Nd}{ii}& 1.14 & $ 0.27$ & 13 &  25.6 & 1.53 &&   3.6 & 0.33\\
   5234.210 & \ion{Nd}{ii}& 0.55 & $-0.51$ & 13 &  21.2 & 1.59 &&   ... &  ...\\
   5249.600 & \ion{Nd}{ii}& 0.98 & $ 0.20$ & 13 &  26.9 & 1.47 &&   ... &  ...\\
   5255.510 & \ion{Nd}{ii}& 0.20 & $-0.67$ & 13 &  25.8 & 1.53 &&   ... &  ...\\
   5311.480 & \ion{Nd}{ii}& 0.99 & $-0.42$ & 13 &   8.3 & 1.44 &&   ... &  ...\\
   5319.820 & \ion{Nd}{ii}& 0.55 & $-0.21$ & 12 &  31.1 & 1.55 &&   3.6 & 0.20\\
   5385.890 & \ion{Nd}{ii}& 0.74 & $-0.82$ & 12 &   7.6 & 1.54 &&   ... &  ...\\
            &             &      & $     $ &    &       &      &&       &     \\
   4244.700 & \ion{Sm}{ii}& 0.28 & $-0.73$ & 14 &  10.1 & 0.93 &&   ... &  ...\\
   4434.320 & \ion{Sm}{ii}& 0.38 & $-0.26$ & 15 &  24.6 & 1.06 &&   4.6 & $-0.04$\\
   4458.520 & \ion{Sm}{ii}& 0.10 & $-0.78$ & 15 &  14.3 & 0.97 &&   ... &  ...\\
   4519.630 & \ion{Sm}{ii}& 0.54 & $-0.43$ & 14 &  14.8 & 1.07 &&   ... &  ...\\
   4523.910 & \ion{Sm}{ii}& 0.43 & $-0.58$ & 14 &  13.7 & 1.06 &&   ... &  ...\\
   4566.210 & \ion{Sm}{ii}& 0.33 & $-0.92$ & 15 &   9.6 & 1.11 &&   ... &  ...\\
   4642.230 & \ion{Sm}{ii}& 0.38 & $-0.52$ & 14 &  12.1 & 0.93 &&   2.4 & $-0.09$\\
   4674.600 & \ion{Sm}{ii}& 0.18 & $-0.56$ & 15 &  15.6 & 0.90 &&   3.7 & $-0.07$\\
   4687.180 & \ion{Sm}{ii}& 0.04 & $-1.17$ & 15 &   9.7 & 1.07 &&   ... &  ...\\
            &             &      & $     $ &    &       &      &&       &     \\
   4129.700 & \ion{Eu}{ii}& 0.00 & $ 0.22$ & 16 &   syn & $-0.25$ &&   syn & $-0.50$\\
   6645.130 & \ion{Eu}{ii}& 1.38 & $ 0.20$ & 16 &   syn & $-0.30:$ &&   syn & $-0.56$:\\
            &             &      & $     $ &    &       &      &&       &     \\
   4191.080 & \ion{Gd}{ii}& 0.43 & $-0.48$ & 17 &   syn & 0.84 &&   syn & 0.08:\\
            &             &      & $     $ &    &       &      &&       &     \\
   4057.810 & \ion{Pb}{i} & 1.32 & $-0.22$ & 18 &   syn & 2.28 &&   syn & $<0.60$\\
\noalign{\smallskip}
\hline
\noalign{\smallskip}
%\multicolumn{10}{l}{  } \\
\multicolumn{10}{l}{References to Table~\ref{Tab:EWandloggf} $-$ (1) \citet{Hibbert_C}; (2) \citet{Gratton94};}\\
\multicolumn{10}{l}{(3) \citet{Hanna_Y}; (4) \citet{Biem81_Zr}; (5) \citet{Ljung_Zr};}\\
\multicolumn{10}{l}{(6) \citet{Cowley83_Zr}; (7) \citet{Davidson_Ba}; (8) \citet{Lawler_La}; }\\
\multicolumn{10}{l}{(9) \citet{Lawler_Ce}; (10) \citet{Palm00_Ce}; (11) \citet{Mashon_Pr};}\\
\multicolumn{10}{l}{(12) \citet{Meggers75_Nd}; (13) \citet{Hartog_Nd}; (14) \citet{Biem89_Sm}; }\\
\multicolumn{10}{l}{(15) \citet{cb62_Sm}; (16) \citet{Mashon_Eu};}\\
\multicolumn{10}{l}{(17) \citet{Hartog_Gd}; (18) \citet{Biem00_Pb}.}\\
\noalign{\smallskip}
\end{longtable}}

\subsection{Line selection and atomic data}
The atomic lines were selected from previous abundance determinations of
the heavy elements of stars, namely \citet{Sneden83}, \citet{Gratton94}, \citet{Sneden96},
Reddy et al. (1997), \citet{Aoki01, aoki02}, \citet{Cowan02}, \citet{Siva04},
and \citet{John04}. The log{\it\,gf} values are primarily adopted from
high-precision laboratory measurements; references are given in Table~\ref{Tab:EWandloggf}.

\subsection{Abundance calculations}
For most elements, the abundances are derived from equivalent widths (EWs) of
unblended lines measured by fitting a Gaussian to the line profile.
For blended lines and/or lines with significant hyperfine structure (HFS)
and/or isotope splitting, the abundances are derived by spectrum synthesis.
The model atmospheres of \citet{Kur93} are used, and local thermodynamic
equilibrium (LTE) is assumed.

\begin{figure}
  \includegraphics[width=0.48\textwidth]{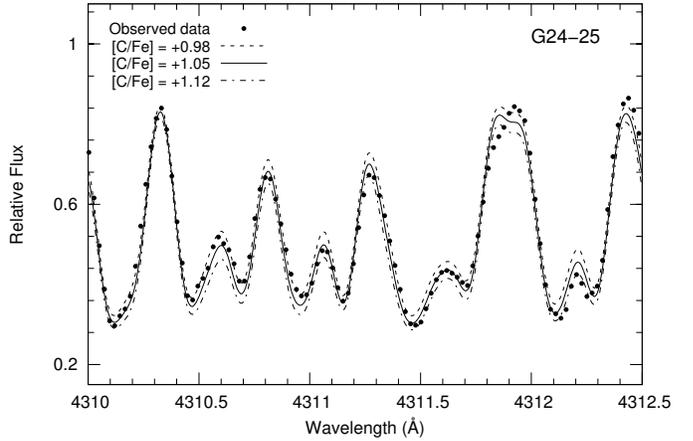}
  \caption{\label{Fig:CH4310} The CH feature near 4310\,\AA .
The observed spectrum of \object{G\,24-25} is shown as bold dots. The solid line shows the
synthetic spectrum corresponding to [C/Fe] = 1.05. The dot-dashed and dashed lines,
corresponding to $\Delta$[C/Fe] = $\pm$0.07, are shown to illustrate the
sensitivity of line strengths to variations in the carbon abundance.}
\end{figure}

The carbon abundance is derived from the $EW$s of three unblended
\ion{C}{i} atomic lines providing [C/Fe] = 1.03 and [C/Fe] = $-$0.39 for \object{G\,24-25} and \object{G\,16-20}, respectively.
Compared to \object{G\,16-20}, \object{G\,24-25} has clear carbon
atomic lines and strong CH and C$_{2}$ lines. From spectrum
synthesis fitting of the CH A-X band of \object{G\,24-25} around 4310\,\AA\
(Fig.~\ref{Fig:CH4310}), we obtained [C/Fe] = 1.05 when
using the CH molecular line data of \citet{Barklem05}.
The abundances derived from the \ion{C}{i} atomic lines and
the CH A-X band are consistent, whereas \citet{Siva04}
reported that carbon abundances derived from \ion{C}{i} lines
are on average about 0.1-0.3 dex higher than those obtained
from the CH lines.

Since the CN 3800\,\AA\ band is not covered in our spectra, we tried
to derive the nitrogen abundance by spectrum synthesis fitting of the
CN 4215\,\AA\ band adopting molecular line data from
\citet{Aoki02a}. The CN lines around 4215\,\AA\ are quite weak
($EW < 5$\,m\AA ), and due to the  rather low $S/N$ ratio in this spectral
region, we can only get an upper limit [N/Fe] $< 0.16$\,dex for \object{G\,24-25}.
In the case of \object{G\,16-20} we get [N/Fe] $< 0.0$\,dex.

The abundances of the light $s$-process elements Sr\footnote{We
do not include the \ion{Sr}{ii} line at 4161.8\,\AA , because the abundance
inferred from that line is 0.3\,dex higher than the average for the other lines
\citep{Sneden96, Aoki01}. We also avoid the \ion{Sr}{ii}
$\lambda 4215.5$ line, which is blended with an \ion{Fe}{i} line.}, Y, and Zr,
as well as the heavy neutron-capture elements Ba, La, Ce, Nd, and Sm are
derived from EW measurements, whereas spectrum synthesis is used for Pr, Eu, Gd, and Pb.
The HFS of the \ion{Ba}{ii} lines at 5853.7, 6141.7, and 6496.9\,\AA\
is not taken into account since it is insignificant according
to \citet{Sneden96} and \citet{Mashon10}. Our Ba abundances are in good agreement
with those in \citet{NS2011}, who also found that the effects of HFS
on the Ba lines are negligible. The \ion{La}{ii} and \ion{Sm}{ii}
lines are quite weak, and we do not include any HFS.

The collisional broadening of lines induced by neutral hydrogen is also considered.
The width cross-sections of the \ion{C}{i}, \ion{Sr}{i}, \ion{Sr}{ii}, and \ion{Ba}{ii}
lines are taken from \citet{Anstee95}, \citet{Barklem97, Barklem00b}, and
\citet{Barklem00a}. For the remaining lines, we follow \citet{Cohen03} and 
adopt the \citet{unsold55} approximation to the van der Waals interaction 
constant enhanced by a factor of two.
For the four strong lines, \ion{Sr}{ii} 4077.7\,\AA\
and the \ion{Ba}{ii} lines at 5853.7, 6141.7, and 6496.9\,\AA ,
the effect on the derived abundances of the uncertainties in
the damping constants is approximately 0.08 dex. However,
most element abundances in our work are based on
weak lines ($EW < 60$\,m\AA), for which the influence on
the results caused by the uncertainties in
the collisional cross-sections is negligible.
Hence, we do not discuss possible errors in
the damping constants in Sect.~\ref{sect:erros}.

\begin{figure}
  \includegraphics[width=0.45\textwidth]{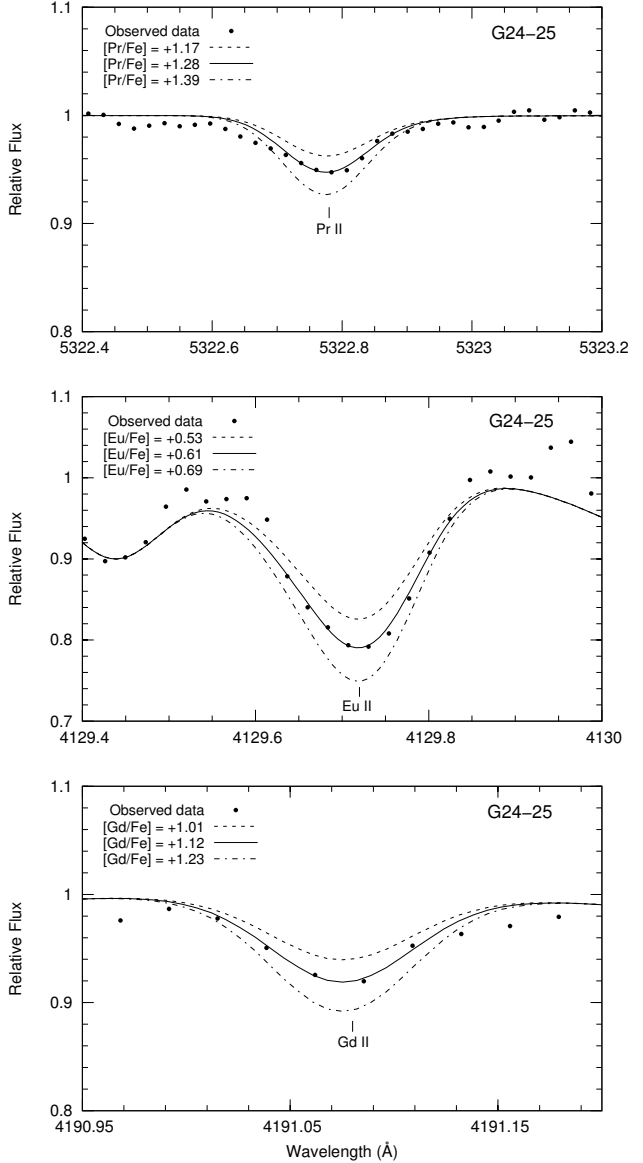}
  \caption{\label{Fig:syn}  Comparison of observed (bold dots) and synthetic
profiles of the spectral lines of the heavy elements Pr (top panel), Eu (middle), and Gd (bottom).
The respective elemental abundances corresponding to the full drawn lines are listed in
Table~\ref{Tab:Abundances}.  Two alternative synthetic spectra are shown to demonstrate the
sensitivity of the line strength to the abundance.}
\end{figure}

The spectrum synthesis fits to the Pr, Eu, and Gd lines are shown in Fig.~\ref{Fig:syn}.
The atomic and hyperfine structure data for the \ion{Pr}{i} 5322.8\,\AA\
and \ion{Eu}{ii} 4129.7\,\AA\ lines are adopted
from \citet{Mashon_Pr} and \citet{Mashon_Eu}, respectively.
The log{\it\,gf} value for the \ion{Gd}{ii} line at 4191.1\,\AA\ is taken from \citet{Hartog_Gd}.

\begin{figure}
  \includegraphics[width=0.45\textwidth]{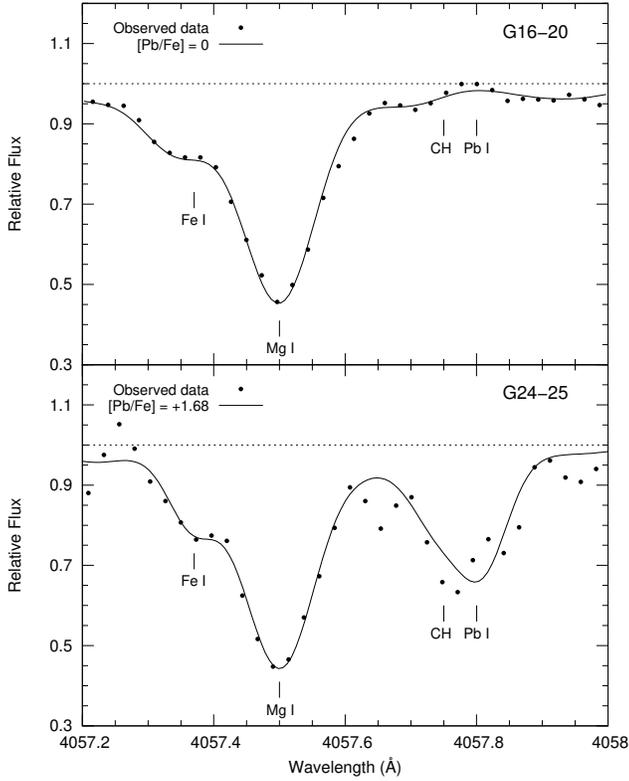}
  \caption{\label{Fig:synPb} Spectrum-synthesis fitting of the \ion{Pb}{I}
$\lambda 4057.81$ line for \object{G\,16-20} and \object{G\,24-25}.}
\end{figure}

\begin{figure}
  \includegraphics[width=0.45\textwidth]{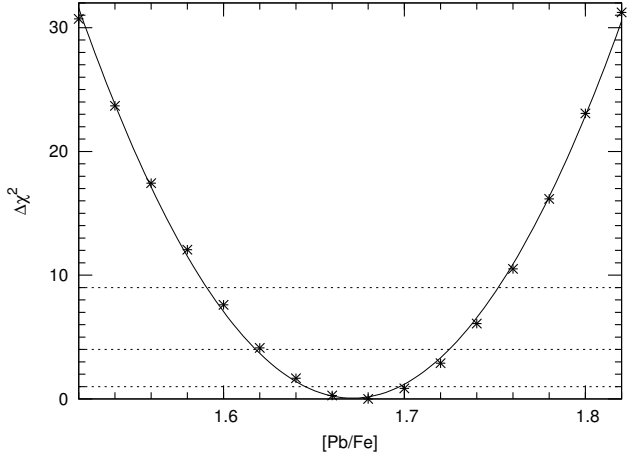}
  \caption{\label{Fig:chisquare} The $\chi^2$ variation in the synthetic fit
to ten observed data points from 4057.69\,\AA\ to 4057.92\,\AA\ applied to the
determination of the lead abundance in \object{G\,24-25}.}
\end{figure}

In the observed spectrum of \object{G\,24-25},
we could detect one lead line, the \ion{Pb}{i} $\lambda 4057.8$,
which is too weak to be seen
in the spectrum of \object{G\,16-20} (Fig.~\ref{Fig:synPb}).
The log{\it\,gf} value is taken from \citet{Biem00_Pb} and the hyperfine data
are taken from \citet{Aoki01}. Compared to \object{G\,16-20}, a more normal neutron-capture star,
it is clear that \object{G\,24-25} exhibits a strong Pb line.
The spectral region from 4057.6\,\AA\ to 4058\,\AA\ does not show a
smooth line profile for \object{G\,24-25}, but considering that
the estimated $S/N$ around the Pb line is
only about 60, we propose that this is probably due to noise.
Thus, a $\chi^2$ fitting of the ten observed data points from
4057.69\,\AA\ to
4057.92\,\AA\ was applied to determine the lead abundance.
The $\chi^2$ was computed as described by \citet{nis99}, i.e.
\begin{displaymath} \ \rm \chi^2=\frac{\sum(O_i-S_i)^2}{\sigma^2},
\end{displaymath}
where $O_i$ is the observed relative flux, $S_i$ the synthetic flux,
and $\sigma$ = $(S/N)^{-1}$ = 1/60. The value of [Pb/Fe] was varied
in steps of 0.02 dex to find the lowest $\chi^2$. The result is a parabolic variation
in $\chi^2$ as shown in Fig.~\ref{Fig:chisquare}. The most probable value
of [Pb/Fe] corresponds to the minimum of $\chi^2$,
and $\Delta\chi^2$ = 1, 4, and 9 (the dashed horizontal lines)
correspond to the 1-,\linebreak 2-, and 3-$\sigma$ confidence
limits of determining exclusively [Pb/Fe].
The spectrum synthesis fit to the Pb line for
the abundance corresponding to the minimum $\chi^2$ ([Pb/Fe] = 1.68)
is shown as the solid line in Fig.~\ref{Fig:synPb}.
A Gaussian broadening function was applied to fit the observed
spectrum in order to take the combined effect
of instrumental, macro-turbulent, and rotational broadening into account.
The nearby spectral lines could be well-fitted in this way leading
to a FWHM of the Gaussian function of $7.8 \pm 1$\,km\,s$^{-1}$.
The corresponding uncertainty in [Pb/Fe] is around 0.04\,dex.

\subsection{Error analysis}
\label{sect:erros}
The uncertainties in the abundances arise from random and systematic errors.
The random errors associated with the errors in the log{\it\,gf} values and
the EW measurements can be estimated from the line-to-line
scatter in the abundances of a given element, based on many lines. The random
error in the mean abundance is ${\sigma}_{\rm EW}/{\sqrt{N}}$, where $N$ is the
number of lines studied. For the elements that have only one line, we use
three times the uncertainty given by \citet{Cay88} by taking into consideration
the error in the continuum rectification.  For instance, \ion{Sr}{i} 4607.34\,\AA\
with $S/N$ = 130 leads to an error of 0.6\,m\AA\ by using the formula (Eq.\ 7)
given by \citet{Cay88}.  We then estimate that the error in the EW measurement is
1.8\,m\AA\ for \ion{Sr}{i} 4607.34\,\AA , which results in an error of 0.06\,dex
in the \ion{Sr}{i} abundance calculation.

A significant contribution to the systematic errors comes from the uncertainties
in the stellar atmospheric parameters adopted in the abundance determination.
This was estimated by varying the model parameters with
$\Delta T_{\mbox{\scriptsize eff}}$ = 100 K,
$\Delta$\,log$g$ = 0.10\,dex, $\Delta$[Fe/H] = 0.1\,dex, and $\Delta \xi$ = 0.30\,km\,s$^{-1}$.
The corresponding abundance uncertainties are listed in Tables~\ref{Tab:Uncertainties1}
and \ref{Tab:Uncertainties2}, and the total uncertainty for each element is estimated by the quadratic sum of the atmospheric and random errors.

An additional systematic error in the derived abundances relative to the
solar abundances is introduced by our LTE assumption and the use of one-dimensional
model atmospheres. These errors tend, however, to cancel for the abundances of
\object{G\,24-25} relative to \object{G\,16-20}, because the two stars have similar
atmospheric parameters and also comparable abundances of Fe and the
$\alpha$-capture elements.

\begin{table}
 \caption{\label{Tab:Uncertainties1} Abundance uncertainties for neutron-capture
   elements in \object{G\,24-25}.}
 \centering
  \begin{tabular}{rcccccc}\hline\hline
    \noalign{\smallskip}
    Elem. & \multicolumn{1}{c}{$\Delta T$} & \multicolumn{1}{c}{$\Delta \log g$} & \multicolumn{1}{c}{$\Delta$ [Fe/H]} & \multicolumn{1}{c}{$\Delta \xi$} & $\frac{\sigma_{\rm EW}}{\sqrt{N}}$ &  $\sigma_{tot}$ \\
      & $+100$~K & $+0.10$ & $+0.10$ & $+0.30$ & & \\
    \hline
    \noalign{\smallskip}
    \ion{C}{i}   & $ -0.06$ & $~~0.03$ & $ -0.01$ & $ -0.01$ & 0.03 & 0.07\\
    \ion{Sr}{i}  & $~~0.08$ & $~~0.00$ & $~~0.01$ & $ -0.02$ & 0.06 & 0.10\\
    \ion{Sr}{ii} & $~~0.06$ & $~~0.01$ & $~~0.03$ & $ -0.02$ & 0.04 & 0.08\\
    \ion{Y}{ii}  & $~~0.05$ & $~~0.03$ & $~~0.00$ & $ -0.06$ & 0.02 & 0.09\\
    \ion{Zr}{i}  & $~~0.08$ & $~~0.00$ & $~~0.00$ & $ -0.01$ & 0.07 & 0.11\\
    \ion{Zr}{ii} & $~~0.04$ & $~~0.04$ & $~~0.00$ & $ -0.04$ & 0.02 & 0.07\\
    \ion{Ba}{ii} & $~~0.09$ & $ -0.02$ & $~~0.01$ & $ -0.10$ & 0.02 & 0.14\\
    \ion{La}{ii} & $~~0.05$ & $~~0.02$ & $~~0.00$ & $ -0.06$ & 0.03 & 0.08\\
    \ion{Ce}{ii} & $~~0.05$ & $~~0.03$ & $~~0.00$ & $ -0.05$ & 0.01 & 0.08\\
    \ion{Pr}{ii} & $~~0.04$ & $~~0.04$ & $ -0.02$ & $ -0.03$ & 0.09 & 0.11\\
    \ion{Nd}{ii} & $~~0.06$ & $~~0.03$ & $~~0.01$ & $ -0.03$ & 0.01 & 0.08\\
    \ion{Sm}{ii} & $~~0.06$ & $~~0.03$ & $~~0.01$ & $ -0.02$ & 0.03 & 0.08\\
    \ion{Eu}{ii} & $~~0.06$ & $~~0.04$ & $ -0.01$ & $~~0.03$ & 0.03 & 0.08\\
    \ion{Gd}{ii} & $~~0.05$ & $~~0.03$ & $ -0.01$ & $ -0.04$ & 0.09 & 0.11\\
    \ion{Pb}{i}  & $~~0.09$ & $~~0.03$ & $ -0.01$ & $~~0.04$ & 0.08 & 0.13\\
    \noalign{\smallskip}
    \hline
  \end{tabular}
\end{table}

\begin{table}
 \caption{\label{Tab:Uncertainties2} Abundance uncertainties for neutron-capture
   elements in \object{G\,16-20}.}
 \centering
  \begin{tabular}{rcccccc}\hline\hline
    \noalign{\smallskip}
    Elem. & \multicolumn{1}{c}{$\Delta T$} & \multicolumn{1}{c}{$\Delta \log g$} & \multicolumn{1}{c}{$\Delta$ [Fe/H]} & \multicolumn{1}{c}{$\Delta \xi$} & $\frac{\sigma_{\rm EW}}{\sqrt{N}}$ &  $\sigma_{tot}$ \\
      & $+100$~K & $+0.10$ & $+0.10$ & $+0.30$ & & \\
    \hline
    \noalign{\smallskip}
    \ion{C}{i}  & $ -0.06$ & 0.04 & $ -0.01$ & $~~0.00$ & 0.09 & 0.12\\
    \ion{Sr}{i} & $~~0.09$ & 0.00 & $~~0.00$ & $ -0.01$ & 0.07 & 0.11\\
    \ion{Sr}{ii}& $~~0.05$ & 0.01 & $~~0.01$ & $ -0.04$ & 0.03 & 0.07\\
    \ion{Y}{ii} & $~~0.04$ & 0.04 & $~~0.01$ & $ -0.01$ & 0.02 & 0.06\\
    \ion{Zr}{ii}& $~~0.04$ & 0.03 & $~~0.01$ & $ -0.02$ & 0.03 & 0.06\\
    \ion{Ba}{ii}& $~~0.06$ & 0.06 & $~~0.01$ & $ -0.08$ & 0.03 & 0.12\\
    \ion{La}{ii}& $~~0.05$ & 0.03 & $~~0.01$ & $ -0.01$ & 0.02 & 0.06\\
    \ion{Ce}{ii}& $~~0.06$ & 0.04 & $~~0.01$ & $~~0.00$ & 0.02 & 0.08\\
    \ion{Nd}{ii}& $~~0.06$ & 0.04 & $~~0.01$ & $~~0.00$ & 0.04 & 0.08\\
    \ion{Sm}{ii}& $~~0.06$ & 0.04 & $~~0.01$ & $ -0.01$ & 0.02 & 0.08\\
    \ion{Eu}{ii}& $~~0.05$ & 0.03 & $ -0.01$ & $~~0.02$ & 0.03 & 0.07\\
    \ion{Gd}{ii}& $~~0.06$ & 0.04 & $ -0.01$ & $ -0.03$ & 0.09 & 0.12\\
    \noalign{\smallskip}
    \hline
  \end{tabular}
\end{table}

\section{Results and discussion}\label{Sect:Discussion}
\subsection{Abundance results}
The abundances for 13 elements are presented in Table~\ref{Tab:Abundances},
along with the number of lines on which they are based,
the adopted solar abundances,
log~$\varepsilon$ = log\,$(N_{\rm X}/N_{\rm H})$ + 12,
and both [X/Fe] and [X/Ba] for each species.
We adopted the standard solar abundances of \citet{Asp09} as
a reference for the various elements. As seen, all of the 13 elements
studied are enhanced in \object{G\,24-25} relative to
both the Sun and the reference star \object{G\,16-20}.

\begin{table*}
 \caption{\label{Tab:Abundances} Abundances of neutron-capture elements of \object{G\,24-25} and \object{G\,16-20}.}
 \centering
 \begin{tabular}{rrrrrrrrrrrrr}\hline\hline
 \noalign{\smallskip}
    Elem. & $Z$ & log~$\varepsilon_{\odot}$ & \multicolumn{4}{c}{\object{G\,24-25}} & & \multicolumn{4}{c}{\object{G\,16-20}}\\
   \cline{4-7}
   \cline{9-12}
   \noalign{\smallskip}
         &     &    & $N_{\rm lines}$ & log~$\varepsilon$ & $\mathrm{[X/Fe]}$ & $\mathrm{[X/Ba]}$ & &$N_{\rm lines}$ & log~$\varepsilon$ & $\mathrm{[X/Fe]}$ & $\mathrm{[X/Ba]}$\\
    \hline
    \noalign{\smallskip}
    \ion{C}{i}   &   6  & 8.39 &  3 &  8.03 & 1.03 & $-0.32$ &&  1 &  6.58: & $-0.39$: & $-0.21$:\\
    \ion{Sr}{i}  &  38  & 2.92 &  1 &  1.96 & 0.44 & $-0.91$ &&  1 &  1.19 & $-0.31$ & $-0.13$\\
    \ion{Sr}{ii} &  38  & 2.92 &  1 &  1.79 & 0.27 & $-1.08$ &&  1 &  1.09 & $-0.41$ & $-0.23$\\
    \ion{Y}{ii}  &  39  & 2.21 &  8 &  1.51 & 0.70 & $-0.65$ &&  7 &  0.59 & $-0.20$ & $-0.02$\\
    \ion{Zr}{i}  &  40  & 2.59 &  1 &  2.14 & 0.95 & $-0.40$ &&... &   ... & $  ...$ &   ...\\
    \ion{Zr}{ii} &  40  & 2.59 &  7 &  2.14 & 0.95 & $-0.40$ &&  3 &  1.40 & $ 0.23$ &  0.38\\
    \ion{Ba}{ii} &  56  & 2.17 &  3 &  2.12 & 1.35 & $ ... $ &&  3 &  0.57 & $-0.18$ &   ...\\
    \ion{La}{ii} &  57  & 1.13 &  5 &  1.28 & 1.55 & $ 0.20$ &&  2 &  0.06 & $ 0.35$ &  0.53\\
    \ion{Ce}{ii} &  58  & 1.58 & 22 &  1.77 & 1.59 & $ 0.24$ &&  6 &  0.31 & $ 0.15$ &  0.33\\
    \ion{Pr}{ii} &  59  & 0.71 &  1 &  0.59 & 1.28 & $-0.07$ &&  1 & $<-0.45$ & $<0.26$ &$<0.44$\\
    \ion{Nd}{ii} &  60  & 1.45 & 14 &  1.52 & 1.47 & $ 0.12$ &&  6 &  0.33 & $ 0.30$ &  0.48\\
    \ion{Sm}{ii} &  62  & 1.01 &  9 &  1.01 & 1.40 & $ 0.05$ &&  5 & $-0.07$ & $ 0.34$ &  0.52\\
    \ion{Eu}{ii} &  63  & 0.52 &  2 & $-0.27$ & 0.61 & $-0.74$ &&  2 & $-0.53$ & $ 0.37$ &  0.55\\
    \ion{Gd}{ii} &  64  & 1.12 &  1 &  0.84 & 1.12 & $-0.23$ &&  1 &  0.08 & $ 0.38$ &  0.56\\
    \ion{Pb}{i}  &  82  & 2.00 &  1 &  2.28 & 1.68 & $ 0.33$ &&  1 & $<0.60$ & $<0.00$ & $<0.18$\\
    \noalign{\smallskip}
    \hline
\end{tabular}
\end{table*}

 \begin{figure}
  \includegraphics[width=0.48\textwidth]{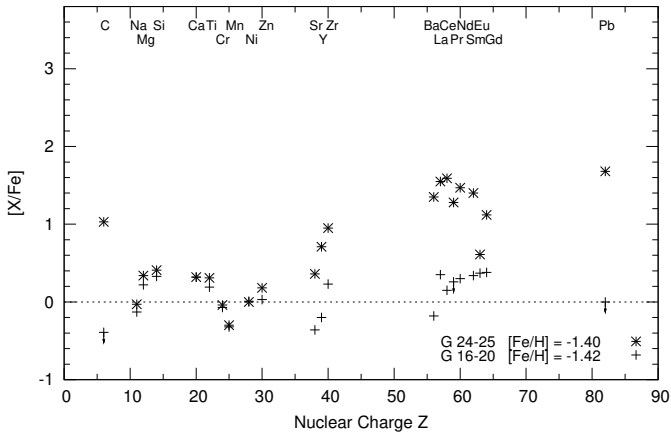}
  \caption{Comparison of abundances in \object{G\,24-25} and \object{G\,16-20}.  \label{Fig:comparetwo}}
 \end{figure}

In Fig.~\ref{Fig:comparetwo}, we present the abundance patterns of
\object{G\,24-25} and
\object{G\,16-20}. The abundances of Na, Mg, Si, Ca, Ti, Cr, Ni,
Mn, Cu, and Zn are adopted from \citet{NS2010,NS2011}.
The two stars have similar behaviors for both the $\alpha$-
and iron-peak elements but show obvious differences for C and
heavy neutron-capture elements.
This can be attributed to the chemical peculiarity of \object{G\,24-25}.
In particular, the second-peak $s$-process elements
(Ba, La, Ce, Pr, Nd, Sm, Gd) are more enhanced
than the first-peak $s$-process elements (Sr, Y, Zr)
in \object{G\,24-25}, which also has a strong carbon enhancement.

Taking into account the most recent research on metal-poor stars,
\citet{BC05} divided metal-poor stars into various subclasses.
According to their definitions, \object{G\,24-25}, which is both
carbon- and barium- enhanced with [C/Fe] $>$ +1.0 and [Ba/Fe] $>$ +1.0 along
with [Ba/Eu] $>$ +0.5, can be classified as CEMP-$s$. However, most CEMP-$s$ stars
found in the literature have [Fe/H] $< -2$, which is more metal-deficient
than \object{G\,24-25} with [Fe/H] = $-1.4$. On the other hand, CH stars
have higher metallicities than CEMP stars and occur in binary systems.
Given that \object{G\,24-25} is a single-lined spectroscopic binary \citep{Lath02},
we suggest that \object{G\,24-25} should be classified as a CH subgiant,
although it has a lower metallicity than the classical CH subgiants belonging to the Galactic disk.
In this context, the C and $s$-process element enhancements are
consistent with mass transfer from a former AGB binary companion \citep{MC90}.

\subsection{Comparing the abundance pattern to model predictions}
\label{Sect:modelling}

To investigate whether the abundance pattern of \object{G\,24-25} can be
explained by mass transfer from a former AGB companion in a binary
system, the observed abundances are compared with the values predicted by
theoretical models. In this comparison, the ratios of the 
abundances of \object{G\,24-25}
to those of the reference star \object{G\,16-20} are used to ensure
that the effects
of Galactic chemical evolution are minimized.

Low-mass AGB stars (1.3 $\la$ $M/M_{\odot}$ $\la$ 3)
are affected by recurrent thermal pulses
in the He shell leading to correlated enhancements of freshly
synthesized $^{12}$C
and $s$-process elements in their photospheres \citep{Stran95, Gallino98}.
In addition, the more massive AGB stars undergo hot bottom burning
where $^{12}$C is converted into $^{14}$N. The minimum mass for this process
is somewhat uncertain, but at the metallicity of \object{G\,24-25},
corresponding to $Z \simeq 0.001$, \citet{Kara07} estimate it to be
around 3.5 $M_{\odot}$ (see their Table 1). Below this mass limit,
an AGB star can have a C-rich atmosphere with no enhancement of nitrogen
as found in the case of \object{G\,24-25} ([N/Fe] $< 0.16$).
Hence, the mass of the AGB star that is responsible for the abundance
anomalies of \object{G\,24-25} is unlikely to be higher than about
3 $M_{\odot}$; a simple consideration of the standard initial mass function implies
that this mass is
more likely to be around 1.5 $M_{\odot}$.

\citet{Bister10} present theoretical predictions of an updated low-mass
AGB stellar nucleosynthesis model at different metallicities. For a model with
a given initial AGB mass and metallicity, they provide the abundances
of all elements from carbon to bismuth in the envelope.
They apply the analysis to CEMP-$s$ stars in
\citet{Bister11} and get a good fit to the observed abundances by
varying the $^{13}$C-pocket profile and dilution factor. Unfortunately,
the calculated model data published in \citet{Bister10, Bister11}
are limited (only two $^{13}$C-pocket cases are available for the mass
and metallicity we need, $M_{ini}^{AGB}$ = 1.5$M_{\odot}$
and [Fe/H] = $-$1.3), and the model prediction does not closely match the
abundances of \object{G\,24-25}.

\begin{figure}
  \includegraphics[width=0.48\textwidth]{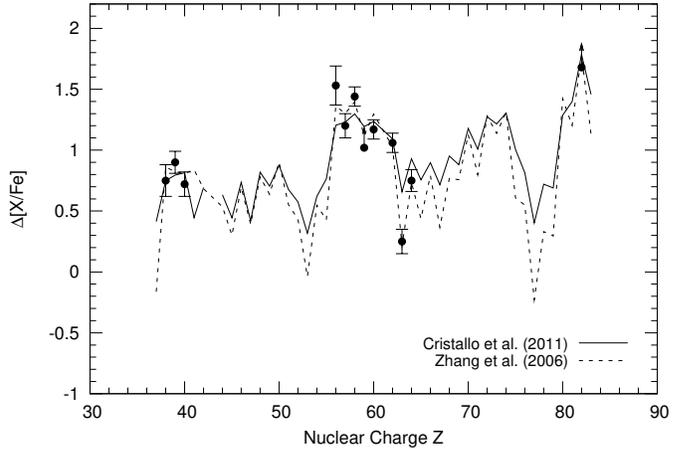}
  \caption{\label{Fig:Atom} The observed abundances of the neutron-capture elements
of \object{G\,24-25} relative to \object{G\,16-20} in comparison
with two AGB model predictions.}
\end{figure}

\citet{Crista09} also present a homogeneous set of calculations of low-mass AGB
models at different metallicities, which they extend to create a
database of AGB nucleosynthesis predictions and yields,
called the FRANEC Repository of Updated Isotopic Tables \& Yield (FRUITY)
database \citep{Crista11}. This database is an online interactive
interface of predictions for the surface composition of AGB stars
undergoing the third dredge-up (TDU) when choosing different combinations of
initial mass and metallicity. The first set of 28 AGB models
available from FRUITY covers the masses 1.5 $\leq$ $M/M_\odot$ $\leq$ 3.0
and metallicities 1$\times$10$^{-3}$ $\leq$ $Z$ $\leq$ 2$\times$10$^{-2}$.
The stellar models of the FRUITY database were computed based on
the updated FRANEC code \citep{Chieffi98}. Carbon-enhanced opacity
tables were adopted to take into account the effects of TDU episodes. The
computation ends when the minimum envelope mass for TDU occurs.
For our purpose, we adopted the AGB model computation from FRUITY
for an initial mass $M = 1.5M_\odot$ and
$Z = 0.001$ (corresponding to [Fe/H] = $-1.3$) and scaled the whole
abundance pattern down to our observational data by a factor of 1.6.
These AGB model predictions for neutron-capture elements from Sr to Pb
is presented by a solid line in Fig.~\ref{Fig:Atom}.
For most elements, the observed abundances are fitted quite well,
but the Eu abundance is significantly overabundant according to the model prediction.

Alternatively, the parametric AGB model for metal-poor stars
from \citet{Zhang06} and \citet{Cui10} has been used to predict the
abundance pattern of \object{G\,24-25}, although this model is
only available for a mass of 3 $M_\odot$ and $Z = 0.0001$. They adopted
the model for metal-poor stars presented by \citet{Aoki01}
after updating many of the neutron-capture rates according to \citet{Bao00}.
This approach is not based on detailed stellar evolution
models, but has been used to successfully explain the abundance patterns of
some very metal-poor stars. There are three parameters in the model:
the neutron exposure per thermal pulse, $\Delta \tau$, the
overlap factor, $r$, and the component coefficient of the $s$-process,
$C_s$. In the case of multiple subsequent exposures, the mean
neutron exposure is given by $\tau_0=-\Delta \tau/{\rm ln}r$.

In their model, the convective He shell and the envelope of the
giant at some time on the AGB, will be overabundant in heavy elements by
factors of $f_{\rm shell}$ and $f_{\rm env,1}$, respectively, with respect to
Solar System abundances normalized to the values of the metallicity, Z.
The approximate relation between $f_{\rm shell}$ and $f_{\rm env,1}$ is
given by Eq.(3) of \citet{Zhang06}, where $\Delta M_{\rm dr}$ is the total
mass dredged up from the He shell into the envelope of the AGB star,
and $M^e_{1}$ is the envelope mass. Given a mass of
3$M_{\odot}$ and Z = 0.0001 for the AGB star, $\Delta
M_{\rm dr}$/$M^e_{1}$ is 1/9 according to \citet{Kara07}.

For a given $s$-process element, the
overabundance factor $f_{\rm env,2}$ in the companion star envelope can
be approximately related to the overabundance factor $f_{\rm env,1}$ by
Eq.(4) of \citet{Zhang06}, where  $\Delta M_{2}$ is the amount
of matter accreted by the companion star from the AGB progenitor, and
$M^e_{2}$ is the envelope mass of the accreting star.  The component
coefficient, $C_s = f_{\rm env,2}/f_{\rm shell}$, is then computed
from Eq.(5) of \citet{Zhang06}.

The closest possible match between the observed and
predicted abundances is obtained for the following values
of the three parameters: overlap factor, $r$\,=\,0.50, neutron
exposure, $\Delta \tau=0.26$\,mb$^{-1}$ (i.e. a mean neutron
exposure $\tau_0=0.375$\,mb$^{-1}$), and $s$-process component
coefficient, $C_s$=0.0012384. The predicted abundance pattern is
given as the dashed line in Fig.~\ref{Fig:Atom}.

To highlight the AGB contribution, Fig.~\ref{Fig:Atom} shows
[X/Fe]$_{\rm G\,24-25}$ minus [X/Fe]$_{\rm G\,16-20}$
as a measure of the abundances of the atmospheric material that has
been accreted by \object{G\,24-25} from its donor star. We note, however,
that the values plotted for [Pr/Fe] and [Pb/Fe] are lower limits to
the difference between \object{G\,24-25} and \object{G\,16-20},
because the Pr and Pb lines could not be detected in
\object{G\,16-20}, hence we have only upper limits to both [Pr/Fe] and [Pb/Fe]
for this reference star. From Fig.~\ref{Fig:Atom}, we can see that the observed
abundance distribution is consistent with the two model predictions,
except that a too high Eu abundance is predicted by the FRUITY
model. The consistency between the observed abundance
distribution and the model predictions indicates that the
overabundances of $s$-process elements in \object{G\,24-25} can be
assumed to originate from mass transfer in a binary system containing material
produced by an AGB companion.

\subsection{Comparison of \object{G\,24-25} with other peculiar stars}
\label{Sect:compare}
\begin{figure}
 \includegraphics[width=0.48\textwidth]{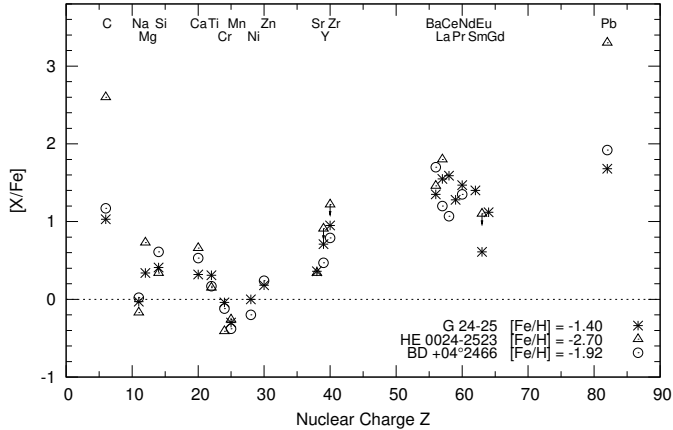}
  \caption{Comparison of \object{G\,24-25} with \object{HE\,0024-2523} (CEMP-$s$ dwarf) and
   \object{BD\,+04$\degr$2466} (CH giant). \label{Fig:comparethree}}
 \end{figure}

It is also interesting to compare the abundance peculiarities
of the neutron-capture elements in \object{G\,24-25} with those of
some well-studied CEMP-$s$ or CH stars.
We chose from the literature stars with as many heavy-element
abundances as possible and atmospheric
parameters close to those of \object{G\,24-25}.
Unfortunately, not one chemically peculiar
star was found to satisfy all these criteria.
Instead, two more metal-poor stars,
\object{HE\,0024-2523} \citep{Luca03} and \object{BD\,+04$\degr$2466} \citep{Pere09}, were considered, both of which have 17 elements
available for comparison and
are members of binary systems.
\object{HE\,0024-2523} is a CEMP-$s$ dwarf star with
[Fe/H] = $-$2.7, and \object{BD\,+04$\degr$2466} is
a classical CH giant with [Fe/H] = $-$1.9. As
seen from Fig. \ref{Fig:comparethree}, \object{G\,24-25}
shows much the same abundance
pattern as the two stars, although \object{G\,24-25} is closer to \object{BD\,+04$\degr$2466},
the classic CH star, than to \object{HE\,0024-2523}.
We note that \object{HE\,0024-2523} was
classified as a lead star with an extreme carbon enhancement
given by [Pb/Fe] = +3.3 and [Pb/Ba] = +1.9.

\subsection{Comparison of [$s$/Fe] versus [Fe/H] for chemically peculiar binary stars}

\begin{figure}
 \includegraphics[width=0.48\textwidth]{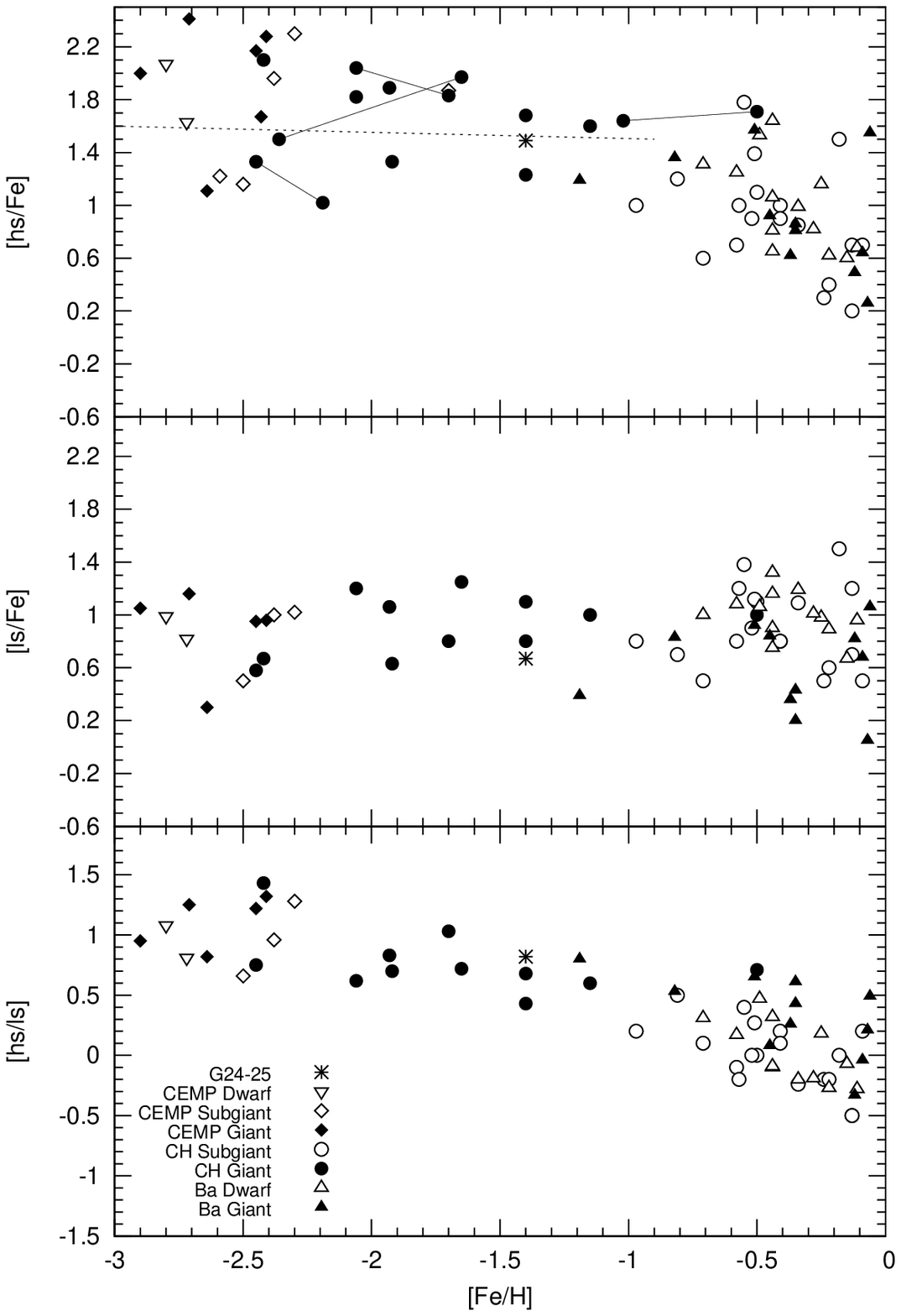}
 \caption{\label{Fig:SvsFe} [$hs$/Fe], [$ls$/Fe] and [$hs/ls$] vs. [Fe/H] are shown in the
three panels, from top to bottom, respectively. Dashed line separates stars belonging
to the high and
the low branches. Data have been collected from the following
references:  CEMP dwarfs - \citet{Luca03},
\citet{Siva04}; CEMP subgiants - \citet{aoki02}, \citet{Cohen03}, \citet{Tsan05}, \citet{Cohen06}, \citet{Thomp08}; CEMP giants - \citet{Aoki01, aoki02}, \citet{Barb05}, \citet{Masser10};
CH subgiants - \citet{LB91}, \citet{Pere03}, \citet{PD11}; CH giants - \citet{Vant92b, Vant92c},
\citet{Kipper96}, \citet{Zac98}, \citet{VanEck01}, \citet{John04}, \citet{Goswa06}, \citet{Pere09},
\citet{Goswa10}, \citet{Masser10};
Ba dwarfs - \citet{AB06}; Ba giants - \citet{AB06}, \citet{Pere09}.
 }
 \end{figure}

Due to the similar formation mechanisms, we have compared \object{G\,24-25} with
other classes of chemically peculiar binary stars, i.e. the CEMP dwarfs, subgiants and giants; the CH subgiants and the classical CH giants; the Ba dwarfs and giants.
The selected binary stars either have derived orbit solutions or exhibit radial velocity variations beyond 3$\sigma$ \citep{Lu87, MC90, PS2001, Aoki03, Cohen03, Luca03, Pour04, Tsan04}.
These various classes are indicated by different symbols in Fig.~\ref{Fig:SvsFe},
where [$hs$/Fe] and [$ls$/Fe] represent the average abundances relative to Fe of the
second-peak elements (Ba, La, Ce, Nd, Sm) and the first-peak $s$-process elements
(Sr, Y, Zr), respectively.  Pr and Gd are not included in the comparison because most works
in the literature have no Pr or Gd measurements.  We do not include the so called
``metal-deficient Ba stars'' (mdBa).  In the past two decades, only five mdBa stars
have been classified based on chemical analyses \citep{LB91, JP01}. Some works
have presented different views of these five ``mdBa'' stars. \object{BD\,+04$\degr$2466}
was confirmed as a CH giant by \citet{Pere09}. \citet{Jori05} considered \object{HD\,104340}
and \object{BD\,+03$\degr$2688} as intrinsic Ba stars, instead of classical Ba stars belonging to a binary system, because they lie on the thermally pulsing AGB sequence and did not show any indication of orbital motion from radial velocity measurements. \citet{Jori05} also doubt the binary nature of \object{HD\,55496} and \object{HD\,206983} since they
had no, or inconclusive, radial velocity data.

For a wide range of [Fe/H] ($-3 <$ [Fe/H] $<$ 0), [$ls$/Fe] has
no dependence on [Fe/H] for all classes of chemically peculiar
binary stars, whereas [$hs$/Fe] shows a decreasing
trend with increasing [Fe/H] albeit with some scatter.
The data of \object{G\,24-25} are consisted with
this trend indicating similar
origins for these stars.  This decreasing trend with [Fe/H]
becomes even more obvious in the [$hs/ls$] versus (vs.) [Fe/H] panel.
In addition, there is a hint that the [$hs/ls$] ([$hs/ls$] = [$hs$/Fe] $-$ [$ls$/Fe]) ratios of giants are systematically higher than
those of subgiants/dwarfs at [Fe/H] $> -0.5$, but further
checks of the differences in the abundance analyses between giants
and subgiants/dwarfs are needed.

\begin{figure}
 \includegraphics[width=0.48\textwidth]{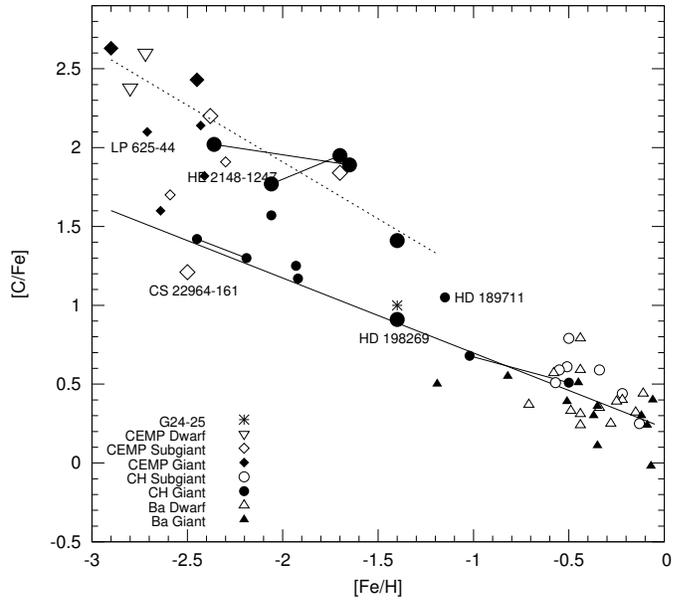}
 \caption{\label{Fig:CvsFe} Comparison of [C/Fe] vs. [Fe/H] for chemically peculiar
binary stars. The big and small sizes of one particular symbol represent
stars with [Pb/Ba] $>$ 1 and [Pb/Ba] $<$ 1, respectively.}
 \end{figure}

According to \citet{LB91} and \citet{Busso01}, [$hs/ls$] is an indicator of the
neutron-exposure in the $s$-process nucleosynthesis: large [$hs/ls$] ratios
correspond to high neutron-exposures.  Therefore, the CEMP, CH, and Ba stars
can be described by
the same scenario in different metallicity ranges where the neutron-exposure in
these stars increases as the metallicity decreases.  In agreement with this trend,
Fig.~\ref{Fig:CvsFe} shows the [C/Fe] ratio as a function of metallicity for the
same objects as in Fig.~\ref{Fig:SvsFe}.  All chemically peculiar binary stars
have a carbon enhancement that increases with decreasing [Fe/H].
The scatter in the
[C/Fe] ratios becomes significant for CEMP stars with [Fe/H] $< -2.0$.  There is
a tendency for the carbon enhancement to diverge into two branches as shown by the
solid (low [C/Fe]) and dashed (high [C/Fe]) lines in
Fig.~\ref{Fig:CvsFe}. Among the high branch, 8 of 11 stars have [Pb/Ba] $>$ 1.0,
but three stars (\object{LP\,625-44}, \object{HE\,2148-1247}, and \object{HD\,189711}) have [Pb/Ba] $<$ 1.0; notably, \object{HD\,189711} has large
uncertainties of 0.5\,dex in [Pb/Fe] \citep{VanEck03}, and \object{HE\,2148-1247}
is a typical CEMP-$r/s$ star \citep{Cohen03}.  On the lower branch,
most stars have [Pb/Ba] $<$ 1.0, except for two stars (\object{CS\,22964-161} and \object{HD\,198269})
for which [Pb/Ba] $>$ 1.0.  According to \citet{Thomp08}, \object{CS\,22964-161} is
a triple system with a double-lined spectroscopic binary and a third component that
might be responsible for its anomalous abundance pattern. The diverging branches
of [C/Fe] versus [Fe/H] are also quite prominent in Fig.~8 of \citet{Pere09},
who make no comment on this. Our star \object{G\,24-25} belongs to the
lower branch.
Careful inspection of the top panel of Fig.~\ref{Fig:SvsFe}
shows evidence of the diverging branches in the [$hs$/Fe] vs. [Fe/H] diagram:
stars located above the
dashed line correspond to the high branch in Fig.~\ref{Fig:CvsFe} (except for HD209621,
a CEMP-$r/s$ star), while those below the dashed line correspond to the low branch.
Such diverging branches could also occur in the [$ls$/Fe] vs. [Fe/H] diagram, but
do not appear to do so, as seen in the middle panel, owing to the smaller effect of the neutron-exposure and large scatter
in the data. This similarity seems to be reasonable in the sense that the enhancements
of C, Pb, $hs$, and probably $ls$ all correspond to the strength of neutron radiation flux
but to a decreasing degree.

\citet{Masser10} completed a holistic study of the CEMP stars. They compiled
abundances from analyses of high resolution spectra of 111 CEMP stars and 21 Ba stars,
including both binary and non-binary stars, and
covering nearly all the chemically peculiar binary stars selected in this work. For four
stars (\object{HD\,26}, \object{HD\,187861}, \object{HD\,196944}, and \object{HD\,224959}),
\citet{Masser10} give updated abundance results.
Comparing the left panel of Fig. 5 in \citet{Masser10} ([Ce/Fe] vs. [Fe/H], where Ce
is a representative $hs$-process element) with the top panel of Fig.~\ref{Fig:SvsFe}
in this work, we can see that [Ce/Fe] shows a significant scatter at all metallicities
and no obvious decreasing trend with increasing [Fe/H], in contrast to our
data for [$hs$/Fe] in Fig. ~\ref{Fig:SvsFe}.
There are two possible reasons for this difference: the first is a statistical
effect where [$hs$/Fe] is based on the average abundance of five $s$-process elements,
which thus has a smaller the scatter; the second is that \citet{Masser10} adopted more stars,
i.e. all CEMP-$s$ stars (the CEMP-low-$s$ included) in their Fig. 5,
but we consider only the $s$-process enhanced binary stars.

\section{Conclusions}\label{Sect:DiscussionConclusions}
On the basis of high resolution and high $S/N$ spectra, the abundances
of carbon and 12 neutron-capture elements, including Pb, are obtained for \object{G\,24-25}.
The C and $s$-process element abundances from the first, second, and third peaks show significant enhancements in this peculiar star with respect to the Sun and also with respect to a reference star \object{G\,16-20}, which has similar atmospheric parameters as \object{G\,24-25}.  The abundance ratios [Pb/Fe] = 1.68 and [Pb/Ba] = 0.33 show that it is not a typical lead star \citep{VanEck01}, which are often found at low metallicity ([Fe/H] $< -2$) to have [Pb/Ba] $> 1.0$.  Owing to its binary signature, high radial velocity, and abundance pattern, we suggest that \object{G\,24-25} is a newly discovered
CH subgiant with a clear detection of the Pb line at $\lambda 4057.8$.

Two simple AGB wind-accretion models have been adopted to predict the theoretical abundances of the $s$-process elements. The comparison of the observed abundances with the model predictions indicates that mass transfer via wind accretion across a binary system from its principal component, a low-mass AGB star, succeeds in reproducing the enhancements of the neutron-capture elements.  When compared with other chemically peculiar binary stars in the literature, we have found that \object{G\,24-25} follows the general trend of [$s$/Fe], [$hs/ls$], and [C/Fe] versus [Fe/H] established for Ba stars, CH stars, and CEMP-$s$ stars implying that there is a similar origin for these objects.

\begin{acknowledgements}
It is a pleasure to thank J. R. Shi and K. F. Tan for valuable discussions
on the determination of lead abundances. We thank B. Zhang and X. J. Shen for
the calculation of and discussions on AGB model predictions.
We thank W. Aoki for
providing the molecular line data for the blue CN band, and an anonymous
referee for helpful comments and suggestions.
This study is supported by the National Natural Science Foundation
of China under grants No. 11073026 and 10821061, the key project of
Chinese Academy of Sciences No. KJCX2-YW-T22, the National Basic
Research Program of China (973 program) No. 2007CB815103/815403.

\end{acknowledgements}

\Online

\end{document}